# Multi-Sensor Conflict Measurement and Information Fusion

Pan Wei, John E. Ball, Derek T. Anderson
Department of Electrical and Computer Engineering, Mississippi State University
Mississippi State, MS USA 39762

## ABSTRACT

In sensing applications where multiple sensors observe the same scene, fusing sensor outputs can provide improved results. However, if some of the sensors are providing lower quality outputs, e.g. when one or more sensors has a poor signal-to-noise ratio (SNR) and therefore provides very noisy data, the fused results can be degraded. In this work, a multi-sensor conflict measure is proposed which estimates multi-sensor conflict by representing each sensor output as interval-valued information and examines the sensor output overlaps on all possible *n*-tuple sensor combinations. The conflict is based on the sizes of the intervals and how many sensors output values lie in these intervals. In this work, conflict is defined in terms of how little the output from multiple sensors overlap. That is, high degrees of overlap mean low sensor conflict, while low degrees of overlap mean high conflict. This work is a preliminary step towards a robust conflict and sensor fusion framework. In addition, a sensor fusion algorithm is proposed based on a weighted sum of sensor outputs, where the weights for each sensor diminish as the conflict measure increases. The proposed methods can be utilized to (1) assess a measure of multi-sensor conflict, and (2) improve sensor output fusion by lessening weighting for sensors with high conflict. Using this measure, a simulated example is given to explain the mechanics of calculating the conflict measure, and stereo camera 3D outputs are analyzed and fused. In the stereo camera case, the sensor output is corrupted by additive impulse noise, DC offset, and Gaussian noise. Impulse noise is common in sensors due to intermittent interference, a DC offset a sensor bias or registration error, and Gaussian noise represents a sensor output with low SNR. The results show that sensor output fusion based on the conflict measure shows improved accuracy over a simple averaging fusion strategy.

**Keywords:** Conflict measure, information fusion, normal measure, fusion

## 1. INTRODUCTION

In classification or clustering, one is predicting or learning “structure” (patterns) in data or identifying anomalies, examples that do not fit the “norm”. In this paper, we discuss an algorithm that measures the regions of overlap, that is, regions where the multiple sensors viewing the same scene have outputs that agree. We call the output of this algorithm the conflict measure (CM). These regions are based solely on the sensor data, and require no other a priori sensor knowledge.

In this paper, we discuss a scenario in which we have no external knowledge of the sensors capabilities (e.g. the signal-to-noise ratios), but instead only have the output data from the sensors. In this work, we assume that all the sensors are viewing the same scene. Since there is no external knowledge about the sensors other than the data itself, one can look at "agreement" and "overlap" in the sensors’ data, since the sensors are all viewing the same scene. Herein, we provide a CM that estimates the conflict between the sensors, in order to understand how well each sensor's outputs agree with the outputs of other sensors viewing the same scene. Furthermore, using the CM for each sensor, one can fuse the sensor data in a more meaningful way. For instance, if we know that a sensors has high conflict with the other sensors, we can assume that the other sensors are probably correct and this sensor is not, and we can de-emphasize the weighting of the conflicting sensor’s outputs when we fuse sensor data together. It is noted that the proposed algorithm is designed to detect conflicts, meaning disagreement between sensors. It is quite another thing to know whether the majority is correct, or if the a single sensor that has a high CM in our method is actually correct. With no knowledge other than the sensor data, one can assume that the majority of non-conflicting sensors has the more correct data. Simply examining the conflicts with the proposed algorithm will not discriminate between these two cases.

*jeball@ece.msstate.edu; phone 1-662-325-4169; www.ece.msstate.edu

By measuring the conflict among sensor data, and using this data, one can potentially facilitate better results in the next processing stage (e.g., sensor output fusion). In the fusion stage, lower weights will be assigned to sources with higher conflict with other sources; on the other hand, higher weights will be assigned to sources with lower CM.

There are numerous sources in the open literature which discuss methods to measure conflict, or in other words, the measure of distance, dissimilarity or divergence[1,2,3,4]. On the other hand, the measure of similarity has been studied in many fields, such as fuzzy measures of agreement[5,6,7,8,9,10] or similarity measures between probability density functions[4]. On some occasions, the measures of distance/dissimilarity/divergence are treated as one minus similarity measure or one over the similarity measure[11]. Herein, the method we propose to calculate conflict has a "one over similarity" relation with similarity measure. However, this relation could be extended to other negative associations.

This paper is organized as follows. Section 2 gives the background of normal measure and fuzzy measure. Section 3 is the proposed algorithm and Section 4 gives several examples of the proposed algorithm. Section 5 contains conclusions and future work.

## 2. BACKGROUND

The proposed conflict measure is a normal measure (NM). The following are definitions of a NM and fuzzy measure (FM)[12]. Let $X = \{x_1,..., x_n\}$ be a set of information sources (e.g., experts, sensors, algorithms, etc.).

**Definition 1. Normal measure (NM)** [13]

Let $(X, \Omega)$ be a measurable space, where $X$ is a set and $\Omega$ is a σ-algebra of $X$. A measure g: $\Omega \rightarrow [0, 1]$ is a NM if there exists a minimal set $A_0$ (e.g., $\varnothing$) and a maximal set $A_m$ (e.g., $X$) in $\Omega$ such that:

1) $g(A_0) = 0$,

2) $g(A_m) = 1$.

Note, often $g(X) = 1$ is for problems like confidence/decision fusion; however, the interval can and has been extended to domains like $[0,R]$, where $R$ is a positive real number.

**Definition 2. Fuzzy measure (FM)** [13,14]

Let $(X, \Omega)$ be a measurable space. A measure g: $\Omega \rightarrow [0,1]$ is a FM if it has the following properties:

1) (Normality) $g(\varnothing) = 0$,

2) $g(X) = 1$,

3) If $A,B \in \Omega$ and $A \subseteq B$, then $g(A) \le g(B)$ .

Similar to the FM, the NM shares the property of boundary conditions. The difference between a FM and NM is that the FM has the condition of monotonicity. For example, if there are three sources, $x_1$, $x_2$, and $x_3$ then the FM

$g(\{x_1,x_2,x_3\})$ is constrained by $g(\{x_1,x_2,x_3\}) \ge g(\{x_1,x_2\})$ , $g(\{x_1,x_2,x_3\}) \ge g(\{x_1,x_3\})$ , and

$g(\{x_1,x_2,x_3\}) \ge g(\{x_2,x_3\})$ . Similarly, $g(\{x_1,x_2\}) \ge g(\{x_1\})$ and $g(\{x_1,x_2\}) \ge g(\{x_2\})$ .

The FM lattice for three sources is shown in Figure 1. The measures are calculated using the proposed algorithm, which is a NM and not a FM. The reason for using a NM is that adding a conflicting source could lessen our measure value; however, a FM cannot due to its monotonicity constraint. The monotonic constraint means that adding another data source (going up the measure) will not lower the fuzzy measure value, which often will not be the case if the aggregating another data source would degrade the results (e.g., when the added sensor output is very noisy compared to the other sensors). Therefore for this work, imposing the monotonicity constraint is not appropriate, so the proposed measure is a NM.

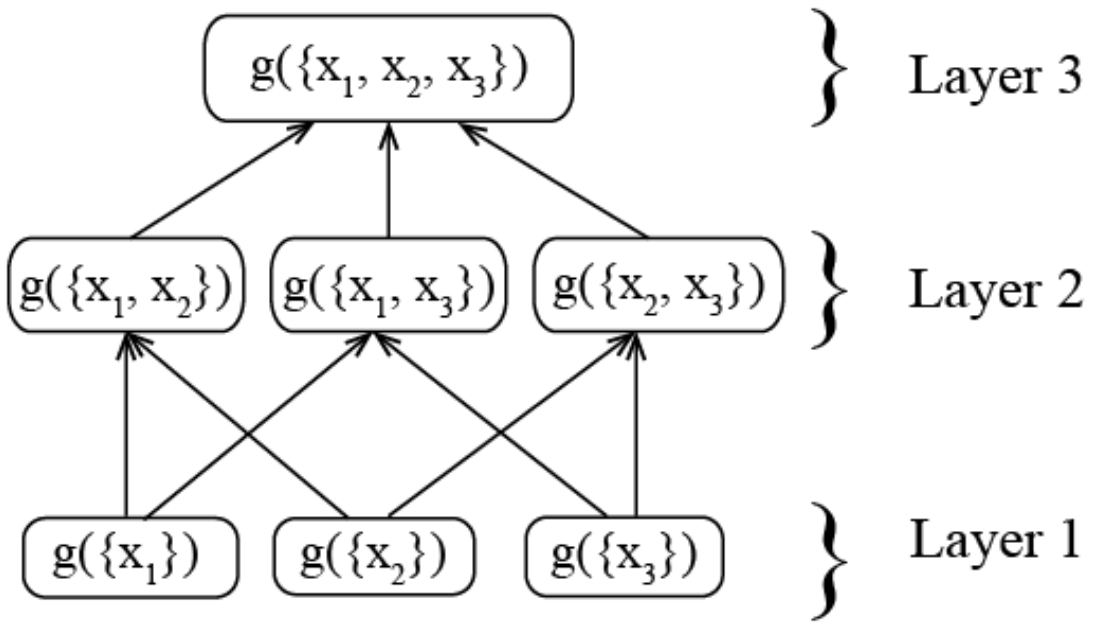

Figure 1. Fuzzy measure lattice for three information sources.

## 3. PROPOSED ALGORITHMS

In this section, the proposed CM is defined and discussed, as well as the algorithms to fuse data based on the proposed CM.

### 3.1 Conflict Measure

The basic idea behind the proposed CM algorithm is to give different weights to different overlapping regions based on how many times these sources' information overlaps in these regions. In a sense, the proposed measure is defining agreement by a combination of the interval sizes and how many sensors are reporting data in the intervals, which is a frequency of overlap approach. The nomenclature and mathematical description of the algorithm is as follows.

Let $X = \{x_1,...,x_n\}$ be a set of sensor sources, for which we do not know a priori the quality of each source. Each source provides data, and subsets of the data can be grouped into interval-valued evidence ($\bar{h}_i$). For example, if sensor $x_1$ has output voltages $v$, where $2 \leq v \leq 3$, then the interval is $\bar{h}_1 = [2, 3]$, and this is called interval-valued evidence. In this case, the interval-valued evidence was based on the minimum and maximum values in the sensor's outputs. This method, however, can lead to an over-inflated interval if there is significant noise or outliers in the data. In this work, we take the data and estimate the mean and standard deviation, and set the interval-valued evidence to the mean plus or minus three standard deviations, which is more robust to outliers.

It is also noted that these intervals are not static since the sensor outputs can change over time. For instance, for a sensor taking data at regular time intervals, one choice is let every $N$ samples constitute a data subset. In the sequel, we will call this data subset a sub-interval. These sub-intervals can be disjoint or overlapping. In this work, we assume overlapping sub-intervals.

Define $A_i$ as the set that contains all $i$-tuple combinations of the sources, where $n$ is the total number of sources. For example, with sources $x_1$, $x_2$ and $x_3$, $A_1 = \{\{x_1\},\{x_2\},\{x_3\}\}$ are all the singletons, $A_2 = \{\{x_1,x_2\},\{x_1,x_3\},\{x_2,x_3\}\}$ are all combinations of pairs of sensors, and $A_3 = \{x_1,x_2,x_3\} = X$ are all triples. Note the sensor order doesn't matter, only the pairings.

Let $E = \{E_1,...,E_{2n}\}$ be the ordered set ($E_i \leq E_{i+1}$) of all interval endpoints from our evidences. Let $\bar{P} = \{\bar{P}_1,...,\bar{P}_{2n-1}\}$ be the set of intervals induced by $E$. Note that each interval requires two endpoints, so there are $2n-1$ intervals since we have $2n$ endpoints. For example, $\bar{P}_1 = [E_1,E_2]$. Let $O(\bar{P}_k)$ be the number of sources with data in the interval $\bar{P}_k$. This is an integral part of our definition of the conflict measure, since more sensors agreeing on an interval means a lower conflict measure, and less sensor agreement means a higher conflict measure.

Let the conflict measure $g^{CF}$ be defined as follows:

$$g^{CF}(A \in A_1) = 0, \quad \forall A \in A_1 \tag{1a}$$

$$g^{CF}\left(A \in A_i\right) = \frac{1}{E_{2n} - E_1} \sum_{k=1}^{2n-1} \left|\bar{P}_k\right| \left(1 - \frac{O\left(\bar{P}_k\right)}{i}\right), \; i \in \{2,3,\cdots,n\} \tag{1b}$$

Eq. (1a) means that for each individual sensor there is no conflict, which is natural, since conflict between sensors really has no meaning for an individual sensor. Eq. (1b) shows that CF is defined here as the sum of sub-interval conflicts over all the sources' measurement range. Each sub-interval contributes to the conflict if some of the sources' measurement range does not cover such sub-interval. The conflict metric is a frequency of overlap approach based on the induced intervals. The overall length of the entire interval $E$ is used as normalizing constant, since the sensor data can live on different intervals.

Once the conflict measures for each $i$-tuple have been calculated using eq. (1a) and (1b), the overall conflict measure for each individual sensor, which is a weighted sum of relative to all the $i$-tuples are calculated as shown in algorithm 1, below. In algorithm 1, for simplicity, the scalar case is shown. The algorithm can easily be extended to vector data. In the vector case, each element, for instance, could be evaluated independently. For example, if the sensor provides 3D data, the algorithm could operate independently on the $x$, $y$, and $z$ coordinates. Since the algorithm processes data in sub-intervals, the sub-interval length must be specified. The length of the evaluation sub-interval will be discussed below.

Algorithm 1. CM estimation algorithm.

**Algorithm to Estimate the Conflict Measure**

**Inputs:** $n$ - the number of sensors

$x_{i,j}$, the $j$-th input from the $i$-th sensor, $i = 1,2,\cdots,n$

$m$ - the number of data points in evaluation sub-interval

**Output:** $g^{CF}\left(A \in A_i\right)$ for $i = 1,2,\cdots,n$

1. For each sensor $i$, $i = 1,2,\cdots,n$
2. For each sub-interval of data with starting index $j$
3. Group the sensor data into a set of sub-interval data: $S_{i,j} = \left\{x_{i,j}, x_{i,j+1}, \cdots, x_{i,j+(m-1)}\right\}$
4. Based on the sub-interval data, estimate the mean value and standard deviation according to

$$\mu_{i,j} = \frac{1}{m}\sum_{k=0}^{m-1} x_{i,j+k} \quad \text{and} \quad \sigma_{i,j} = \sqrt{\frac{1}{m}\sum_{k=0}^{m-1}\left(x_{i,j+k} - \mu_i\right)^2}$$

5. Set the interval-valued evidence $\bar{h}_{i,j} = \left[\mu_{i,j} - 3\sigma_{i,j}, \mu_{i,j} + 3\sigma_{i,j}\right]$
6. Calculate $g^{CF}\left(A \in A_i\right)$ for each element $A \in A_i$ according to eq. (1a) and (1b)
7. End for
8. End for

### 3.2 Fusion based on the Conflict Measure

To fuse the data, it is desired to weight the data such that the weight of a sensor diminishes as the CF increases. Let $w_k$ be the weight of source $k$ . One reasonable approach is to perform the weight calculation for source $j$ is as follows:

$$w_j = \frac{1/SumC_j}{\sum_{i=1}^{n}(1/SumC_i)} , \tag{2}$$

where

$$SumC_j = \sum_{k=1}^{n}\frac{1}{k}\left(\sum_{A\in A_k} g^{CF}(A)\right) , \tag{3}$$

is the weighted sum of the CF values across the lattice. For example, if there are $n = 4$ sensors, for sensor number 1,

$$\begin{aligned} SumC_1 &= \frac{1}{1}g^{CF}(\{x_1\}) \\ &+\frac{1}{2}\left[g^{CF}(\{x_1,x_2\})+g^{CF}(\{x_1,x_3\})+g^{CF}(\{x_1,x_4\})\right] \\ &+\frac{1}{3}\left[g^{CF}(\{x_1,x_2,x_3\})+g^{CF}(\{x_1,x_3,x_4\})+g^{CF}(\{x_2,x_3,x_4\})\right] \\ &+\frac{1}{4}\left[g^{CF}(\{x_1,x_2,x_3,x_4\})\right] \end{aligned}$$

In this manner, the weights produced by eq. (2) will sum to one, and the increased values of CF for a given sensor will lower the weights for that sensor, and thus diminish the effects of sensors that are exhibiting high conflict. Once the weights are calculated, the fused data interval can be calculated as follows:

$$FusedLeftEnd = LeftEnd_1 \times w_1 + LeftEnd_2 \times w_2 + LeftEnd_3 \times w_3 + LeftEnd_4 \times w_4 \tag{4a}$$

$$FusedRightEnd = RightEnd_1 \times w_1 + RightEnd_2 \times w_2 + RightEnd_3 \times w_3 + RightEnd_4 \times w_4 \tag{4b}$$

Finally, the data can be fused using a weighted sum based on the weights calculated in eq. (2) as follows, where the $j$-th data from sensor $i$ is $x_{i,j}$ . For a stereo camera sensor that returns 3D data, one can fuse the $x$, $y$ and $z$ data independently according to algorithm 2.

$$FusedResult_j = \sum_{i=1}^{n} x_{i,j} w_i \tag{5}$$

Algorithm 2. Algorithm to assess the conflict measure for 3D (*x,y,z*) data and fuse the results.

**Inputs:** $\{(x_{i,j}, y_{i,j}, z_{i,j})\}$ *j*-th tracking points from *i*-th sensor

$n$ - the number of sensors

$m$ - the number of data points in the conflict measure sub-interval

Output: $g^{CF}$ NM for *x*, *y* and *z* data

1. Do these steps for each *x* value
2. For $i = 1, 2, \cdots, n$
3. Calculate $g^{CF}(A \in A_i)$ for the *x* data according to algorithm 1
4. End for
5. For $i = 1, 2, \cdots, n$
6. Calculate the weights according to eq. (2) and (3)
7. End
8. Calculate the fused interval using eq. (4a) and (4b)
9. Fuse the data according to the weights using eq. (5)
10. Repeat for each *y* and *z* value

In algorithm 2, the size of the conflict measure sub interval is provided as in input. This size will depend on many factors, and in the section below a reasonable choice of the sub-interval size based on the data is given, based on examination of several different sub-interval sizes.

# 4. EXAMPLES

In this section, a simple numeric example is provided to illustrate the calculations involved. In addition, three stereo camera sensor 3D tracking examples are also provided.

## 4.1 Numeric Example

A synthetic example is given to illustrate the calculation of the CF. For the interested reader, more synthetic examples can be found in [15]. For simplicity, we will assume we are processing only one sub-interval of sensor data. Suppose there are four interval-valued evidences, $\bar{h}_1 = [0, 10]$, $\bar{h}_2 = [2, 8]$, $\bar{h}_3 = [3, 7]$, and $\bar{h}_4 = [4, 6]$. The four evidences are shown in Figure 2(a), while the lattice of CF conflict measures is shown in Figure 2(b). For the calculation of $g^{CF}(\{x_1, x_2, x_3\})$, as shown in Figure 3, intervals [0, 2] and [8, 10] have one source, sensor 1. Intervals [2, 3] and [7, 8] have two overlapping sources which are from sensors 1 and 2. Interval [3, 7] has all three overlapping sources. The CF is computed according to eq. (1b) as follows:

$$g^{CF}(\{x_1, x_2, x_3\})$$
$$= (|0-2| \times \frac{3-1}{3} + |10-8| \times \frac{3-1}{3} + |2-3| \times \frac{3-2}{3} + |8-7| \times \frac{3-2}{3} + |3-7| \times \frac{3-3}{3})/(|10-0|)$$
$$\approx 0.3333$$

The results in Figure 2(b) show that CF is similar to what one might expect in terms of conflict. For example, the algorithm gives higher conflict measure to the set of sources $\{x_1, x_2, x_4\}$ than to the set of sources $\{x_1, x_2, x_3\}$. This is because there is more overlap to sources 1 and 2 from source 3 than from source 4.

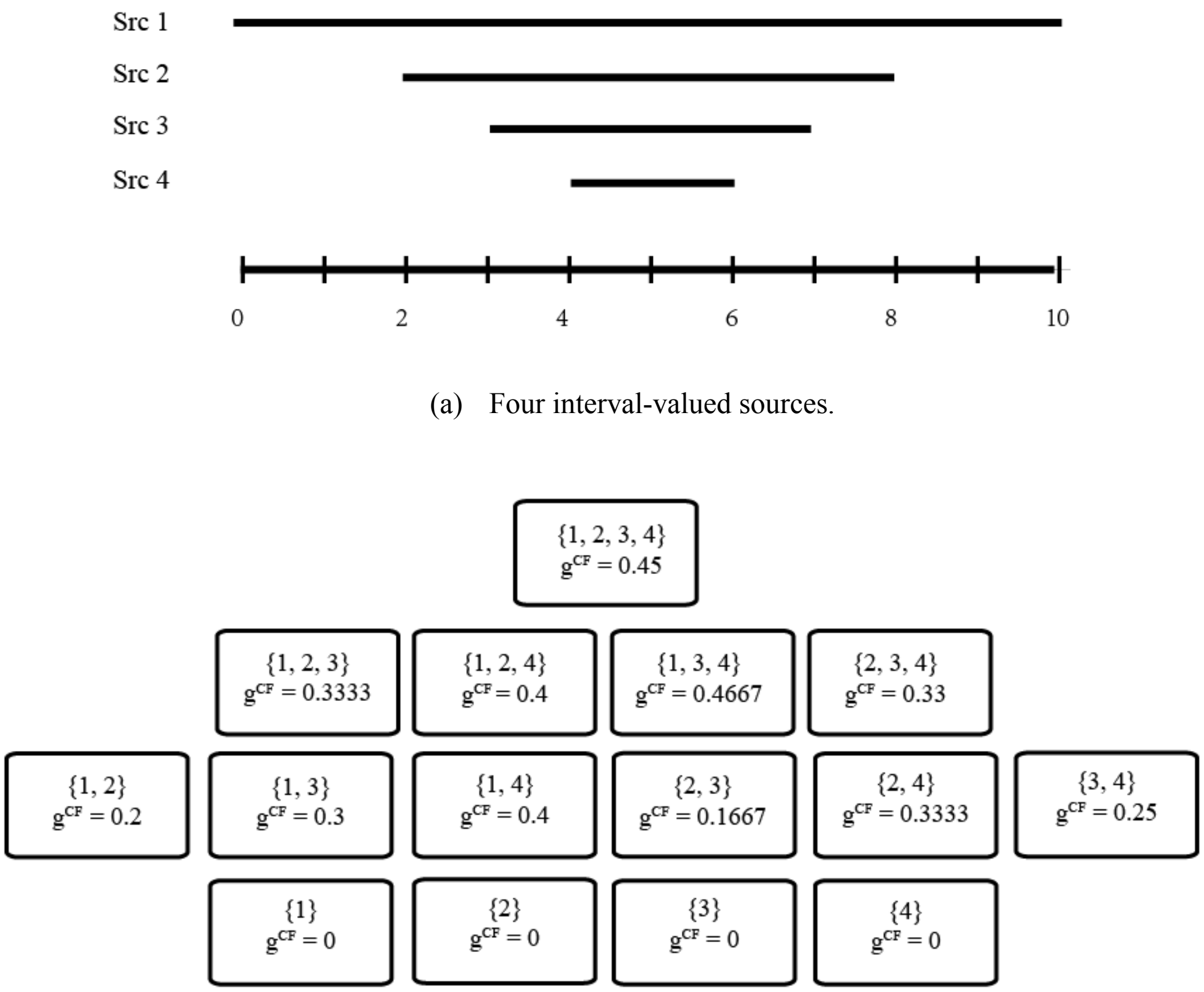

(a) Four interval-valued sources.

(b) Lattice of CF values.

Figure 2. Synthetic Example for calculating CF. (a) Four interval values sources. (b) Lattice of CF values.

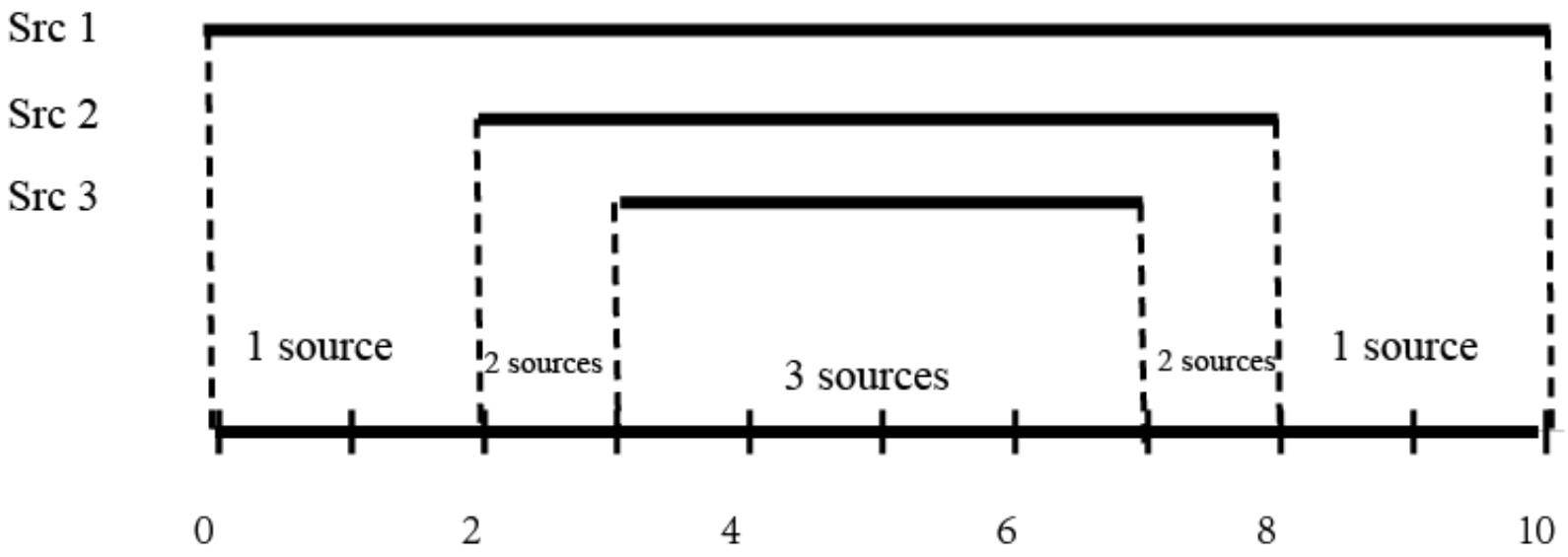

Figure 3. An example showing how to calculate $g^{CF}$ ({*x1, x2, x3*}) of three sources.

To calculate the fusion weights of each source, we first calculate the sum of conflict measures for each source by adding the conflict measures that contain each particular source with a weight. Let $SumC_k$ be the overall sensor CM for source *k,* where $k$=1,2,...,$n$. For example, if there were $n$=4 sensors, then

$$SumC_1 = \frac{1}{1} g^{CF}\left(\{x_1\}\right) + \frac{1}{2}\left[g^{CF}\left(\{x_1,x_2\}\right) + g^{CF}\left(\{x_1,x_3\}\right) + g^{CF}\left(\{x_1,x_4\}\right)\right] + \frac{1}{3}\left[g^{CF}\left(\{x_1,x_2,x_3\}\right) + g^{CF}\left(\{x_1,x_3,x_4\}\right) + g^{CF}\left(\{x_2,x_3,x_4\}\right)\right] + \frac{1}{4}\left[g^{CF}\left(\{x_1,x_2,x_3,x_4\}\right)\right]$$

In this example, $SumC_1 = 0.9625$, $SumC_2 = 0.8180$, $SumC_3 = 0.8486$, $SumC_4 = 1.0042$.

In the proposed method, it is desired that the weight assigned to each source have a negative correlation with the conflict measure. One can choose a function that negatively correlates the input and output, and then normalize it to make the sum of weights from all sources equal to one. In this example, we choose the simplest "one over" functions (e.g., $SumC^{-1}$). There are also other methods such as $SumC^{-n}$, where $n$ could be any positive integer, and other possibilities.

In our example, the weight for sensor 1 is

$$w_1 = \frac{1/SumC_1}{\sum_{i=1}^{n} 1/SumC_i} = \frac{1/0.9625}{1/0.9625 + 1/0.8180 + 1/0.8486 + 1/1.00425} \approx 0.2342$$

Similarly, $w_2 = 0.2756$, $w_3 = 0.2657$, and $w_4 = 0.2245$.

To fuse the three intervals together, one can use eq. (3a) and (3b) as follows:

$$FusedLeftEnd = LeftEnd_1 \times w_1 + LeftEnd_2 \times w_2 + LeftEnd_3 \times w_3 + LeftEnd_4 \times w_4$$
$$= 0 \times 0.2342 + 2 \times 0.2756 + 3 \times 0.2657 + 4 \times 0.2245 = 2.2462$$

The right end could also be calculated in a similar way, which evaluates to 7.7538. The fused result is show in Figure 4.

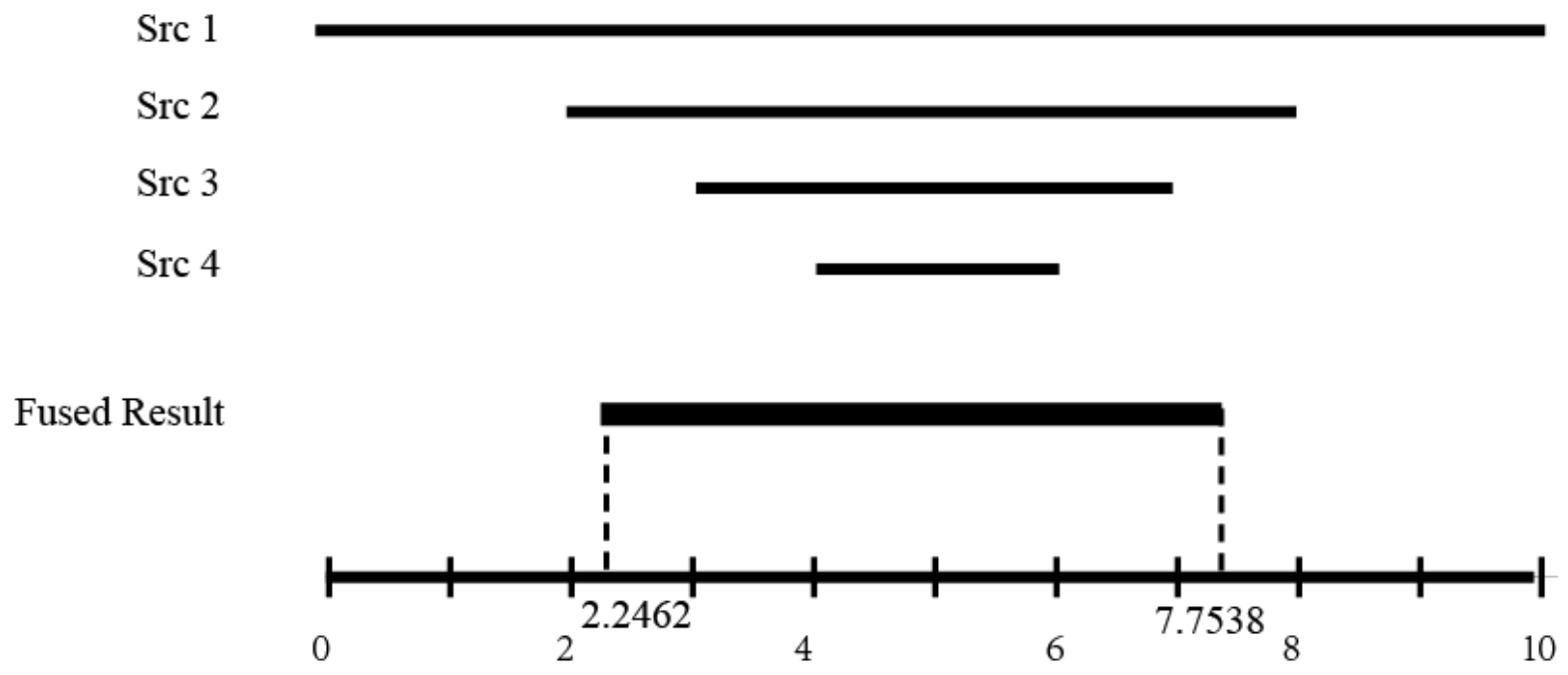

Figure 4. Fused result of four sources using the CF-based weights.

### 4.2 Stereo Camera Experimental Examples

The experimental setup is shown in Figure 5. Four pairs of stereo cameras provide four independent sources of tracking information in 3D space. In this experiment, the stereo cameras are all placed at nearby locations viewing the same object. The cameras were calibrated using the checkerboard poster shown in the left of Figure 5. One stereo camera was chosen as the reference, and all other cameras images were adjusted to the measurement frame of the reference stereo camera[16.] An object was tracked in time by the cameras, and synthetic perturbations were added to stereo camera 2 in the form of additive impulse noise, additive DC offset (bias), and additive Gaussian noise. These examples were chosen to represent some common problems encountered in multi-sensor data. Impulse noise is common in sensors due to intermittent interference, a DC offset a sensor bias or registration error, and Gaussian noise represents a sensor output with low SNR.

The calculation is similar to the synthetic example in 4.1, and the only difference is that instead of using the function $SumC^{-1}$, $SumC^{-3}$ is employed, as it has a better result since it largely diminishes the effect of noise.

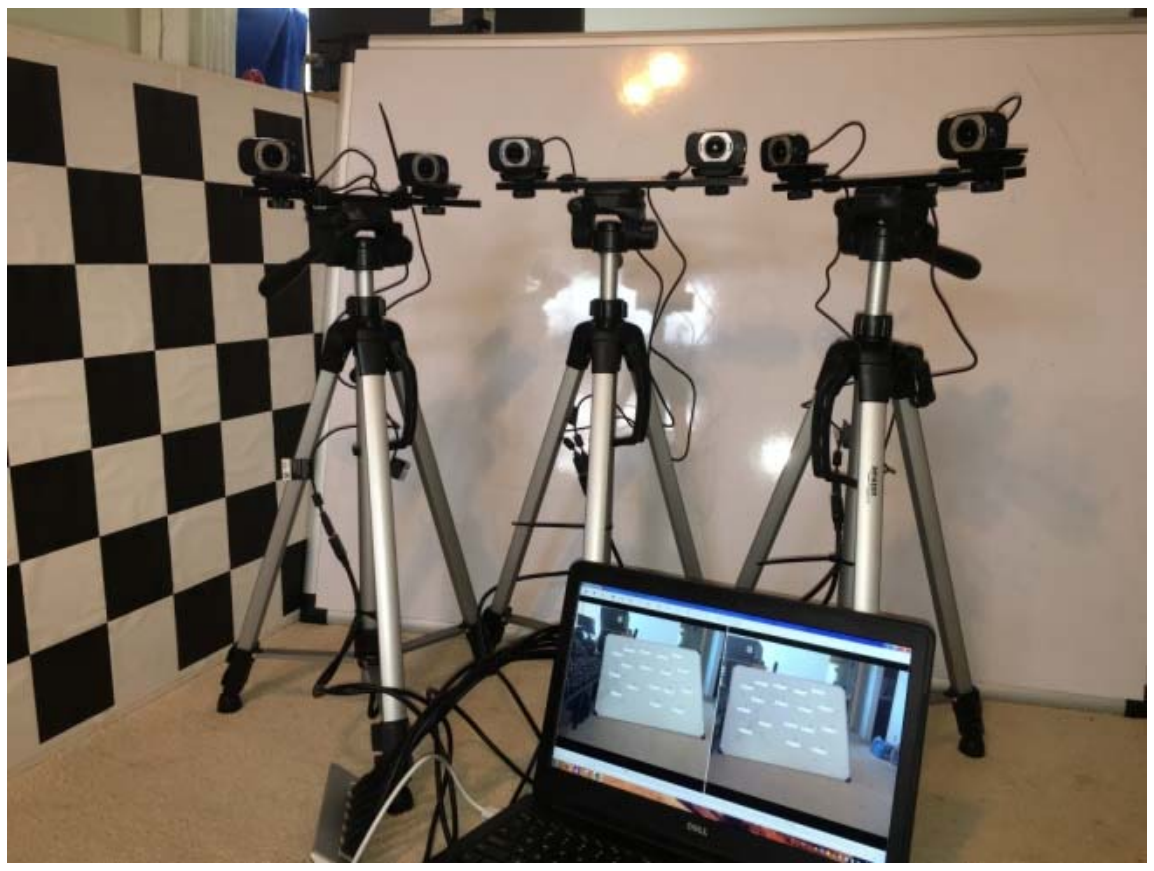

Figure 5. Experimental setup showing the calibration board (to the left), and three of the four stereo camera pairs.

In this experiment, each stereo camera takes snapshots at a fixed time interval. It is assumed that each stereo camera is basically synchronized with respect to time (unsynchronized sources can be considered for future work). In the proposed method, a larger time sub-interval (which is an integer multiple of the sampling time) is chosen to evaluate the sensor measurements using the CF. In our experiments, it was noticed that with the change of the number of points in our sensor evaluation interval, the mean and standard deviation of the conflict measure also change. For example, Figure 6 shows the changing of mean and standard deviation of conflict measure using source 1 and source 4 data. After examination of the different sub-interval lengths, it was empirically discovered for our example case that around interval 10 all the curves change from a large slope down to a smaller slope down. In this case, the sub-interval size was chosen to be 10 times the sample size (that is, ten data points per interval) in the following three experiments. This value is in part due to the inherent 3D motion of the tracked object, as well as the dynamic nature of the conflict measure. Different scenarios may have different curves, and this effect will be studied in future work.

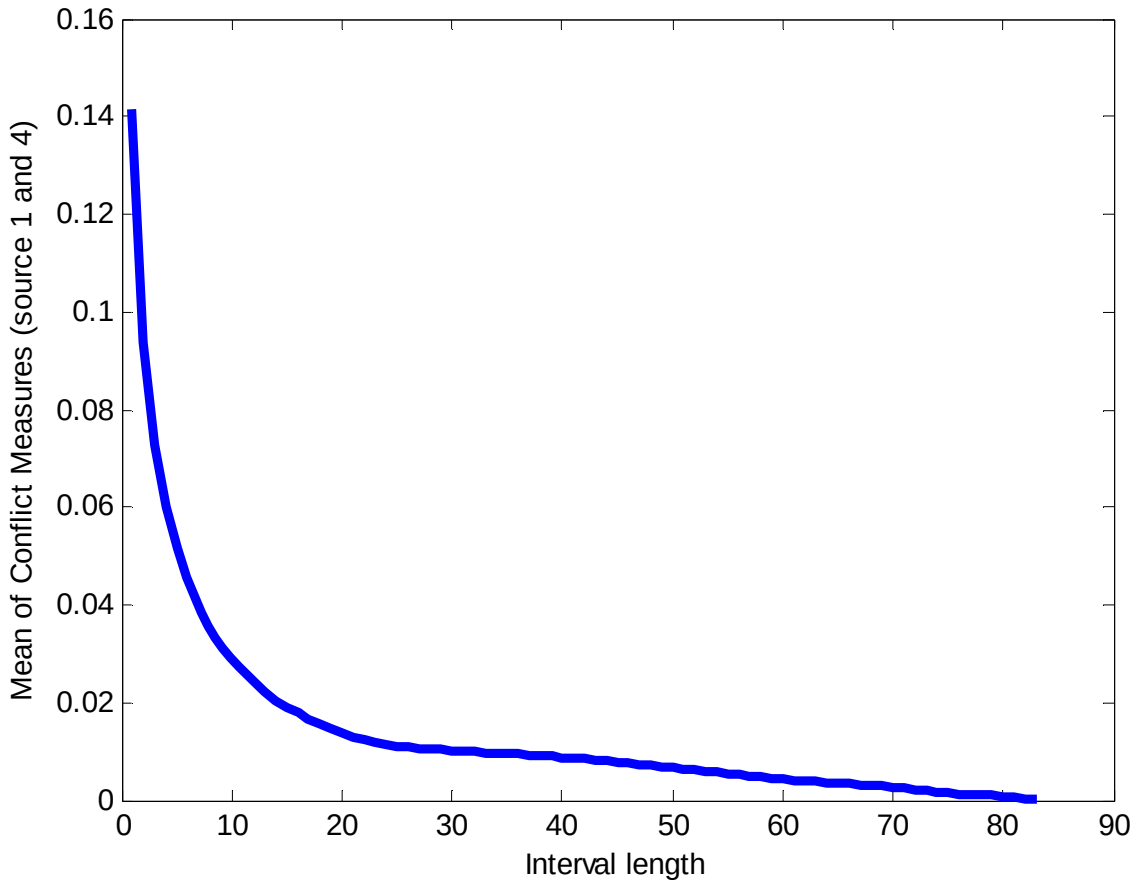

(a)

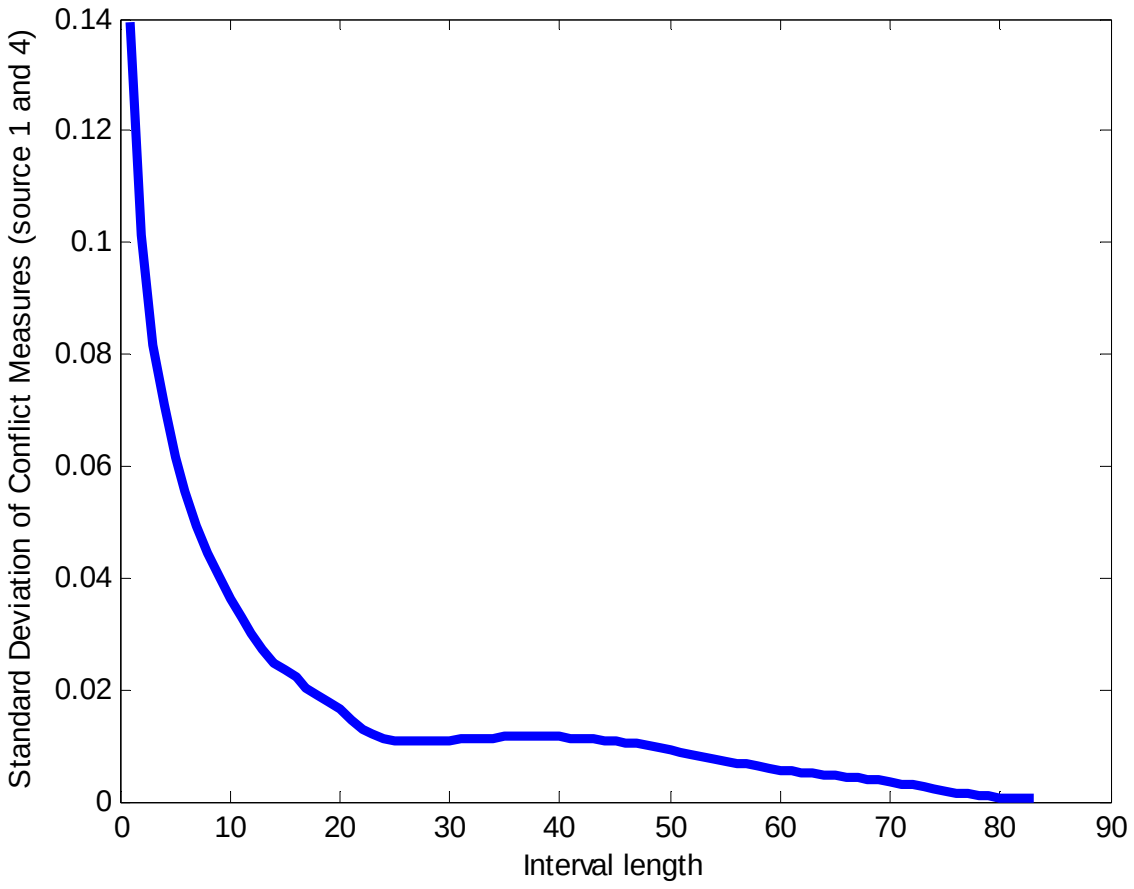

(b)

Figure 6. Mean and standard deviation of conflict measures (from source 1 and source 4) versus the overall interval length, where the interval length is an integer representing the number of samples, which is the CF sub-interval length in Algorithm 1.

The following three examples are based on different noise types which are artificially added to one of the sources in order for us to have a controlled amount of perturbation to the sensor outputs. The three types of noise added are impulse noise, a DC offset, and Gaussian noise. The impulsive noise represents some random interference. This type of noise can play havoc with sensor fusion algorithms due to the outlier nature of the corrupted data. The DC offset represent a misaligned source or a source that has some constant bias. Finally, the Gaussian noise represents a scenario with a poor sensor or a scenario where one sensor has low signal-to-noise ratios. In all of the experiments, sensor 1 and sensor 2 data were unaltered, while the noise was added to sensor three.

**Example 1. Impulse noise**

In this example, five constant impulsive noise of magnitude 100 are added to source 2's $z$ values, and this noise source is treated as source 3, which is shown in Figure 7(a) and (b). The data is color coded as indicated in the legend for Figure 7(a). The result after fusion is shown in Figure 7(d). In the result, we can see that the impact of the impulsive noise is reduced.

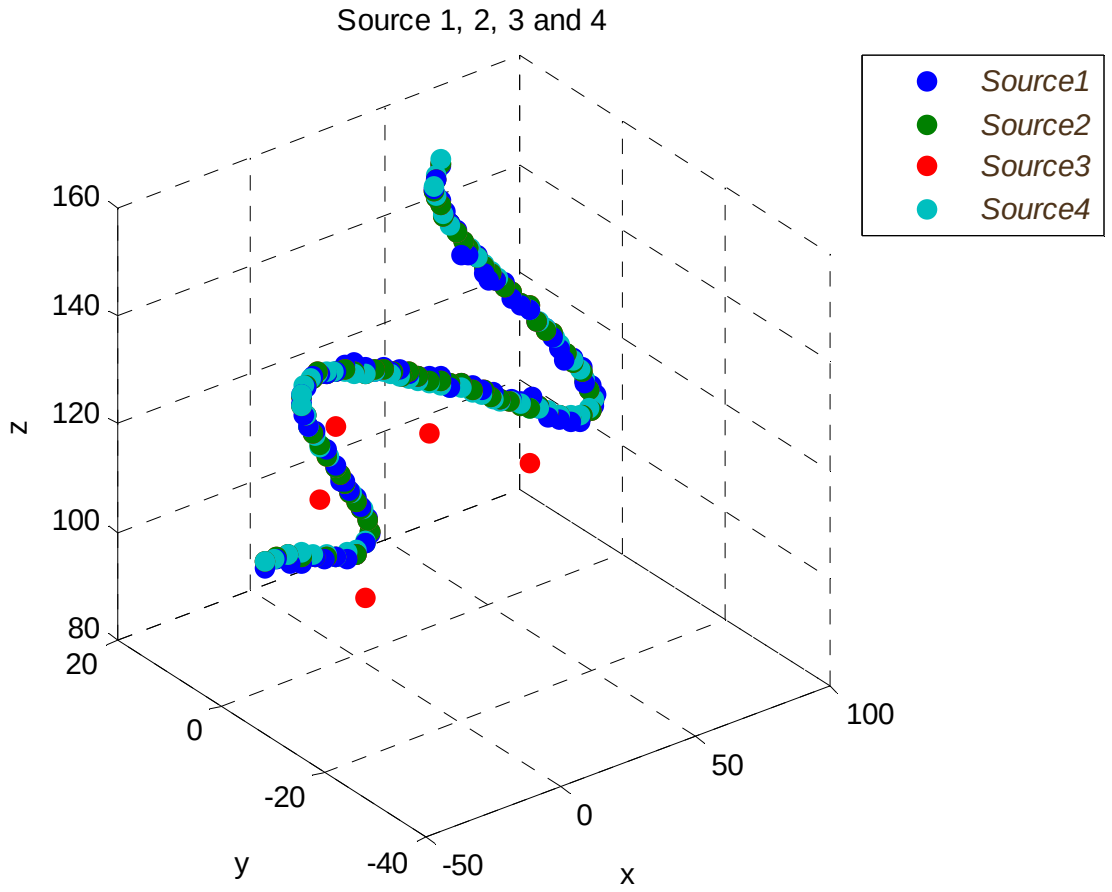

(a) Sensor outputs for all four sensors

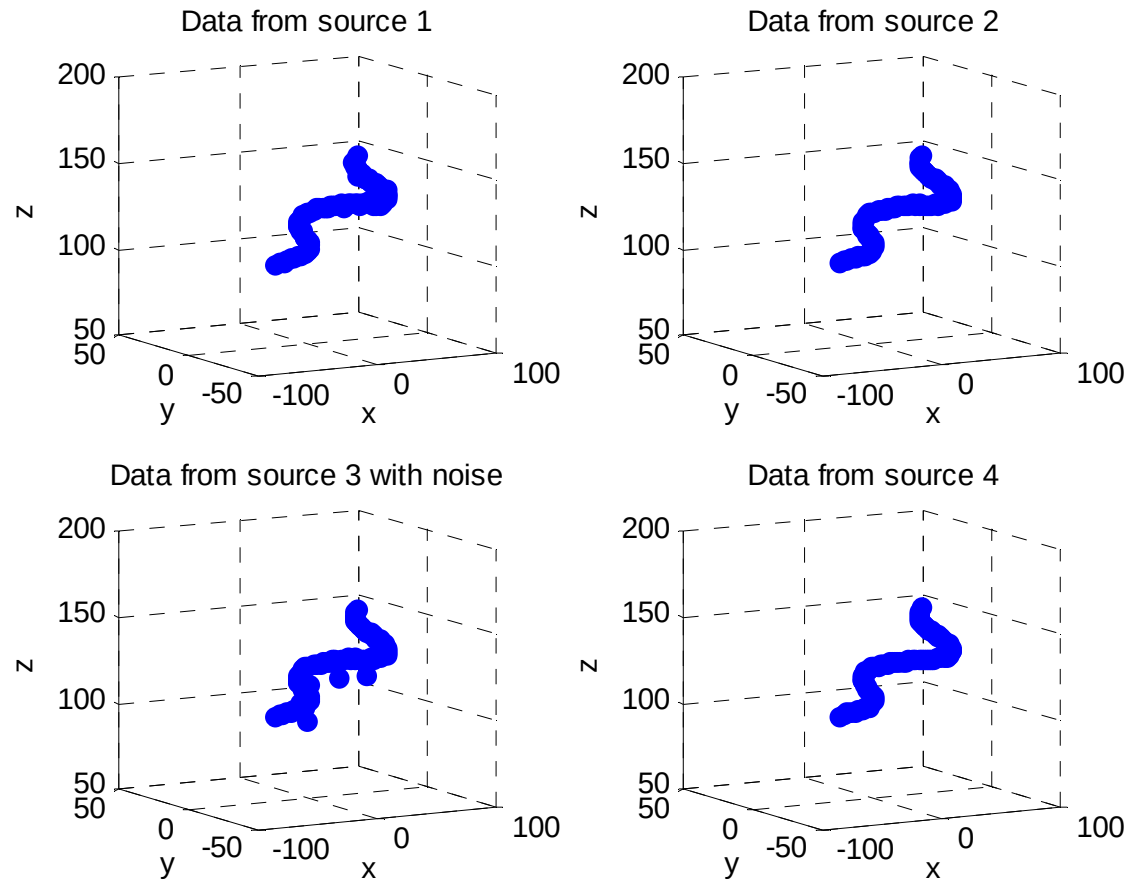

(b) Sensor outputs for each sensor

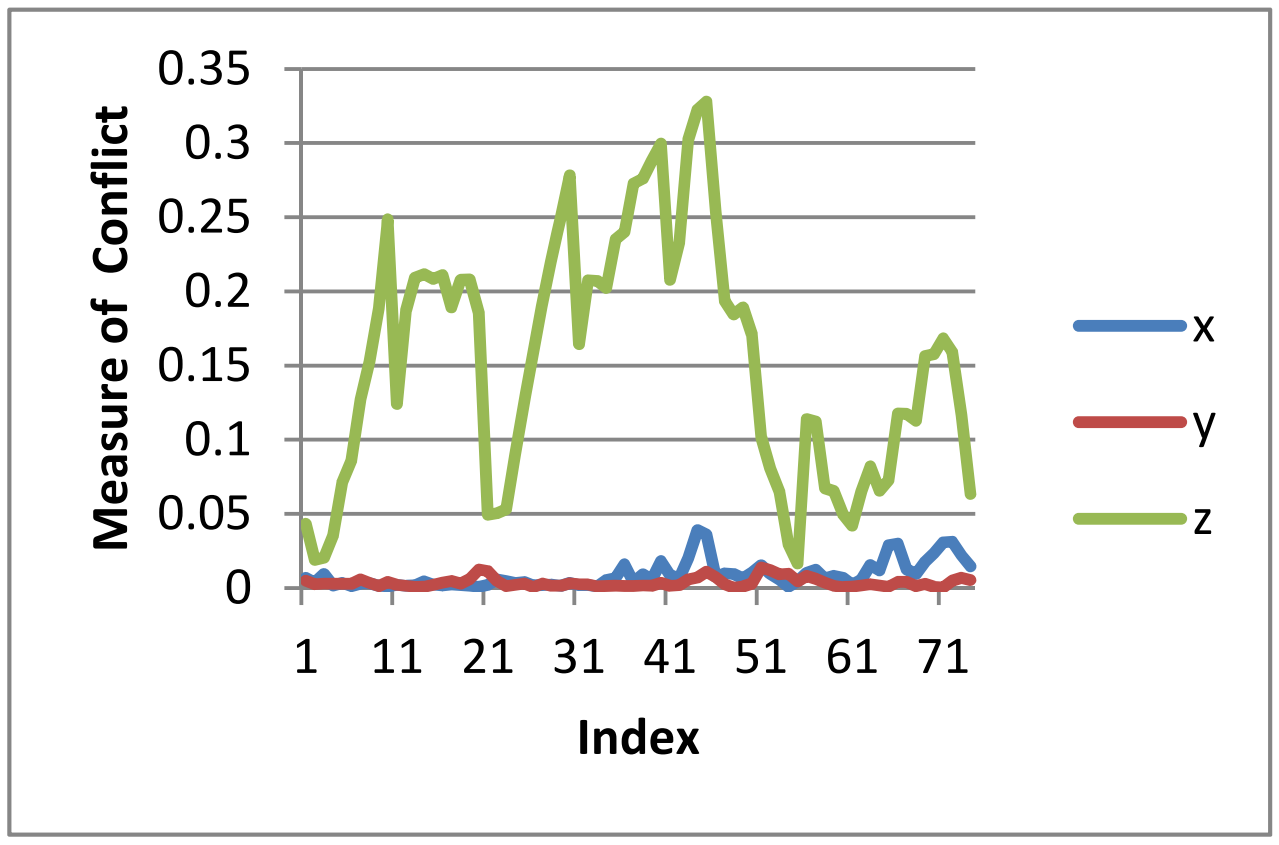

(c) Measure of Conflict of source 1 and source 3

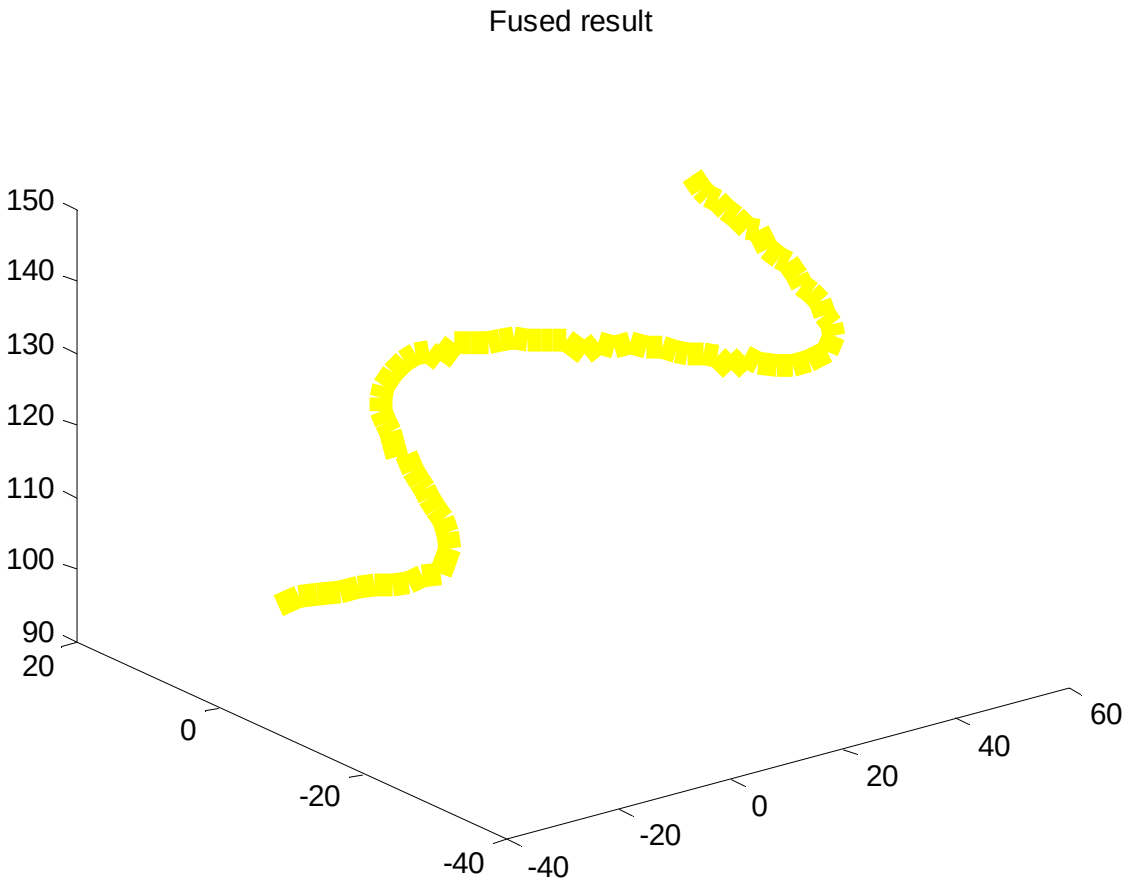

(d) Fused result

Figure 7. (a) Sensor outputs for the four sources, with added impulse noise. (b) Sensor outputs for each source plotted individually. (c) Conflict measure between sources 1 and 3 plotted versus the sub-interval length. (d) Fused result.

**Example 2. DC offset**

In this example, a DC or bias level (-1000) is added to source 2's *z* value, and this noise source is treated as source 3, as shown in Figure 8 (a) and (b). This could be a common problem for multiple sensors where there is a misregistration error or some sort of bias in a distance estimation. Every *z* value of source 3 is lower than the other sources. Figure 8(e) is the result using average operator for comparison. The fused result in Figure 8(d) shows that the result largely reduces the impact of the source with DC offset.

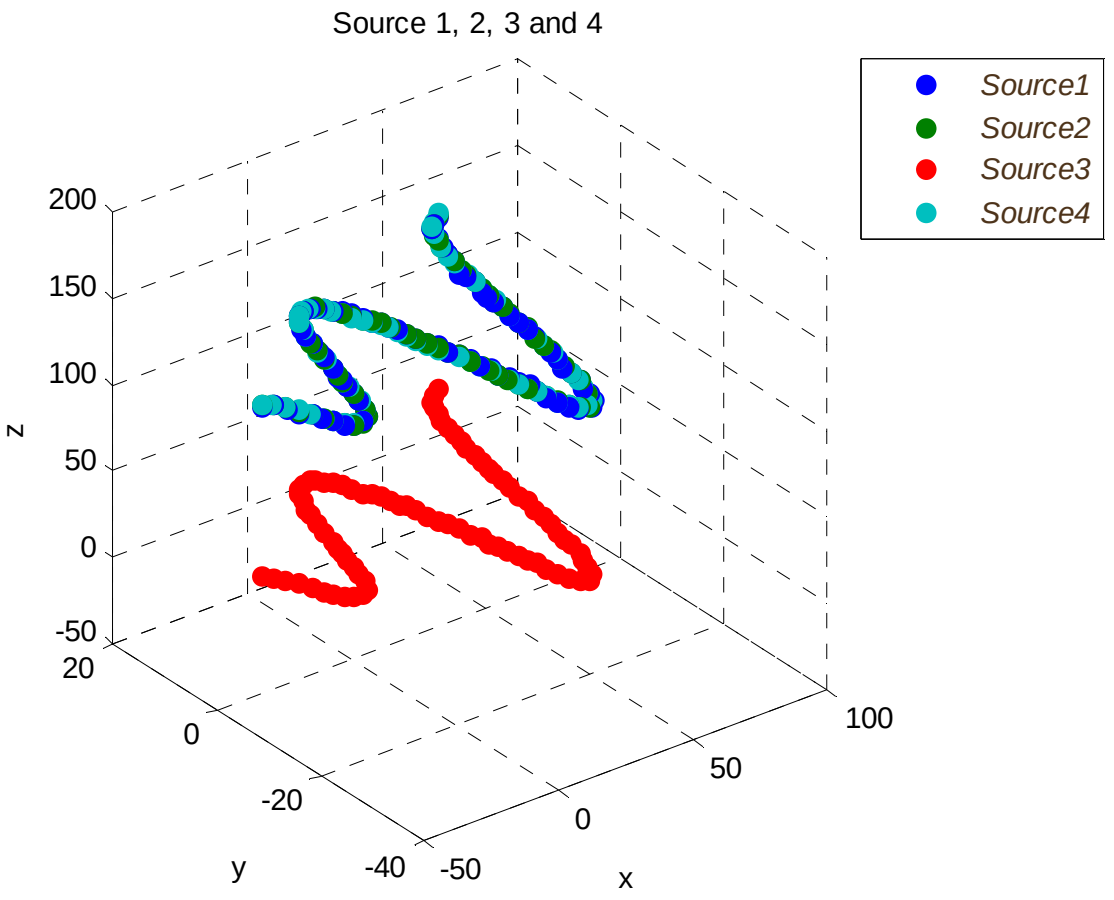

(a) Sensor outputs for all four sensors

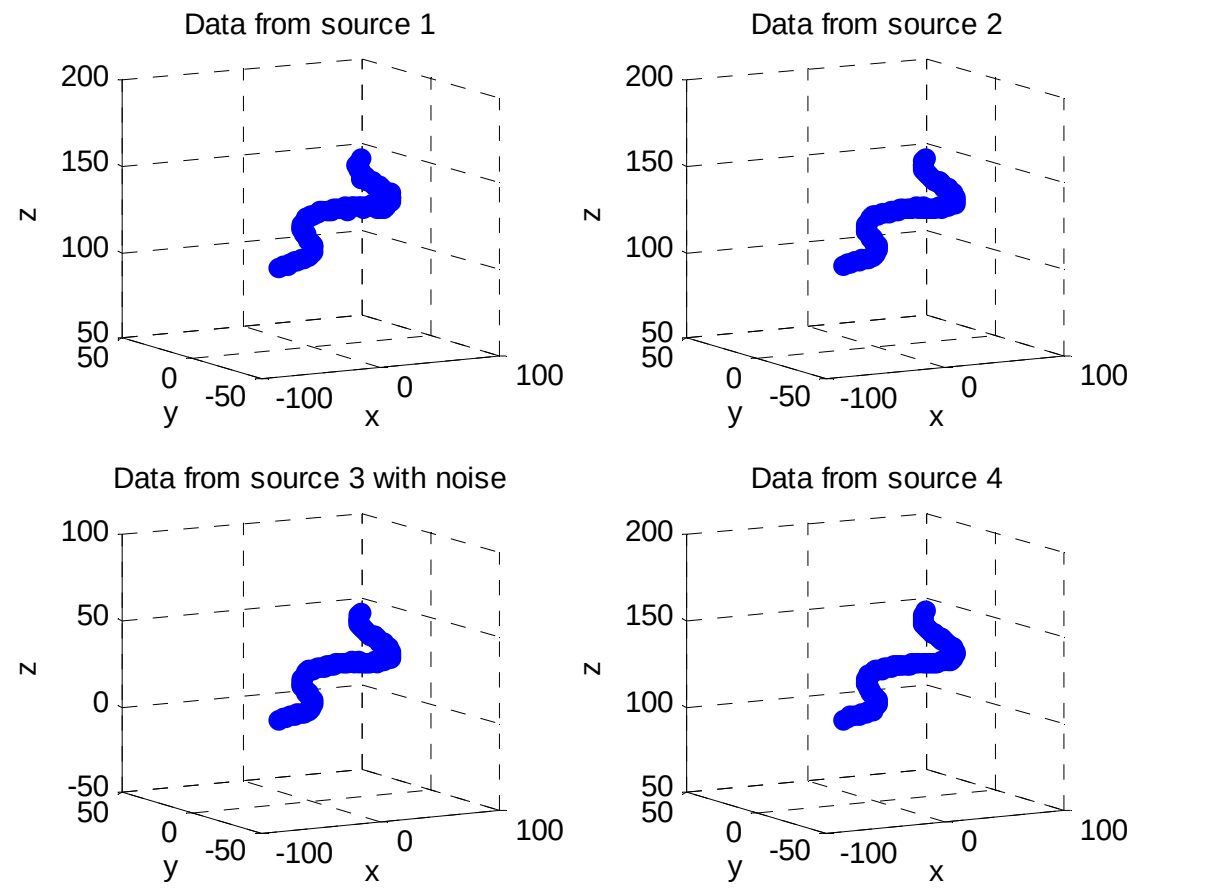

(b) Sensor outputs for each sensor

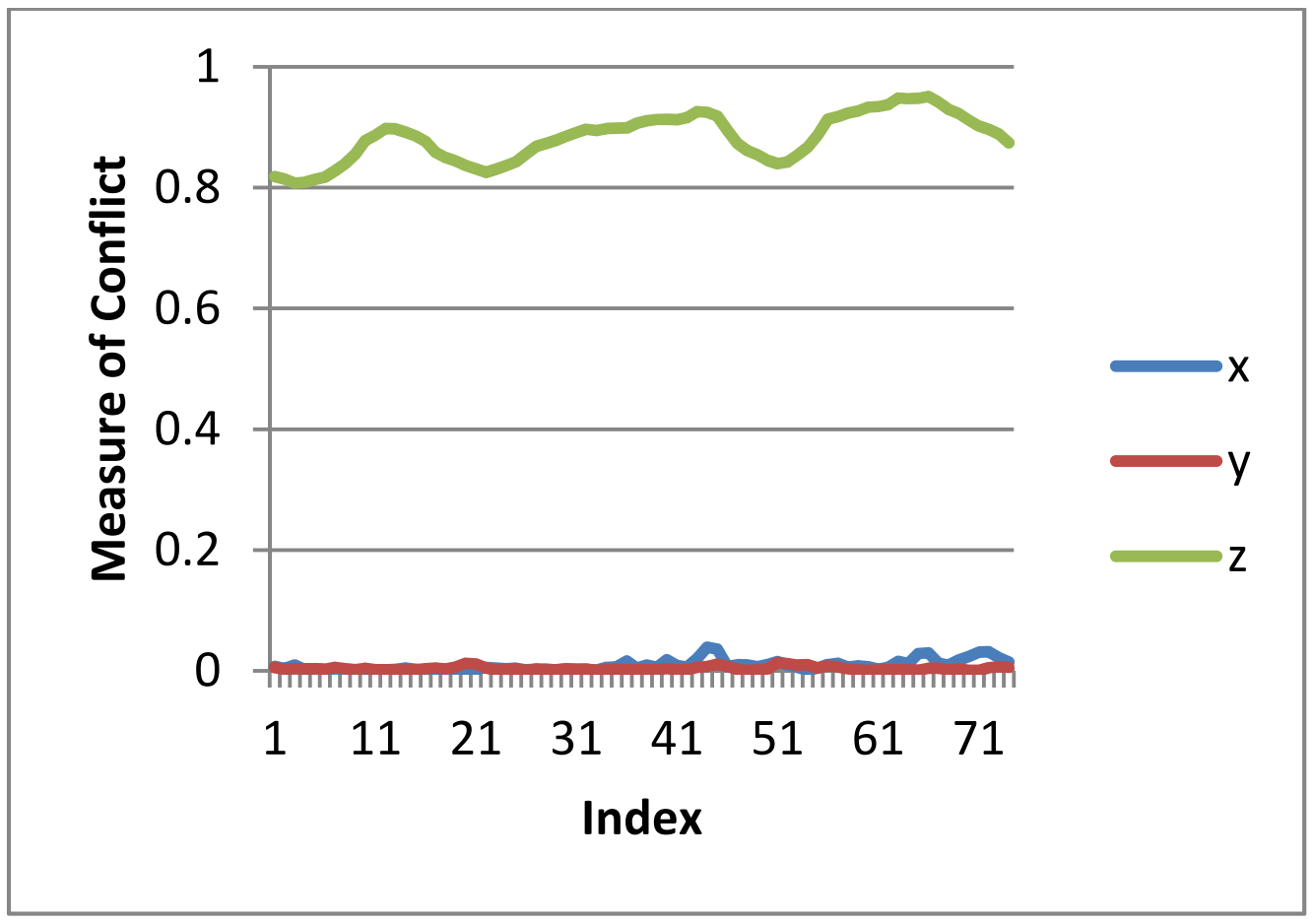

(c)Measure of Conflict of sources 1 and 3

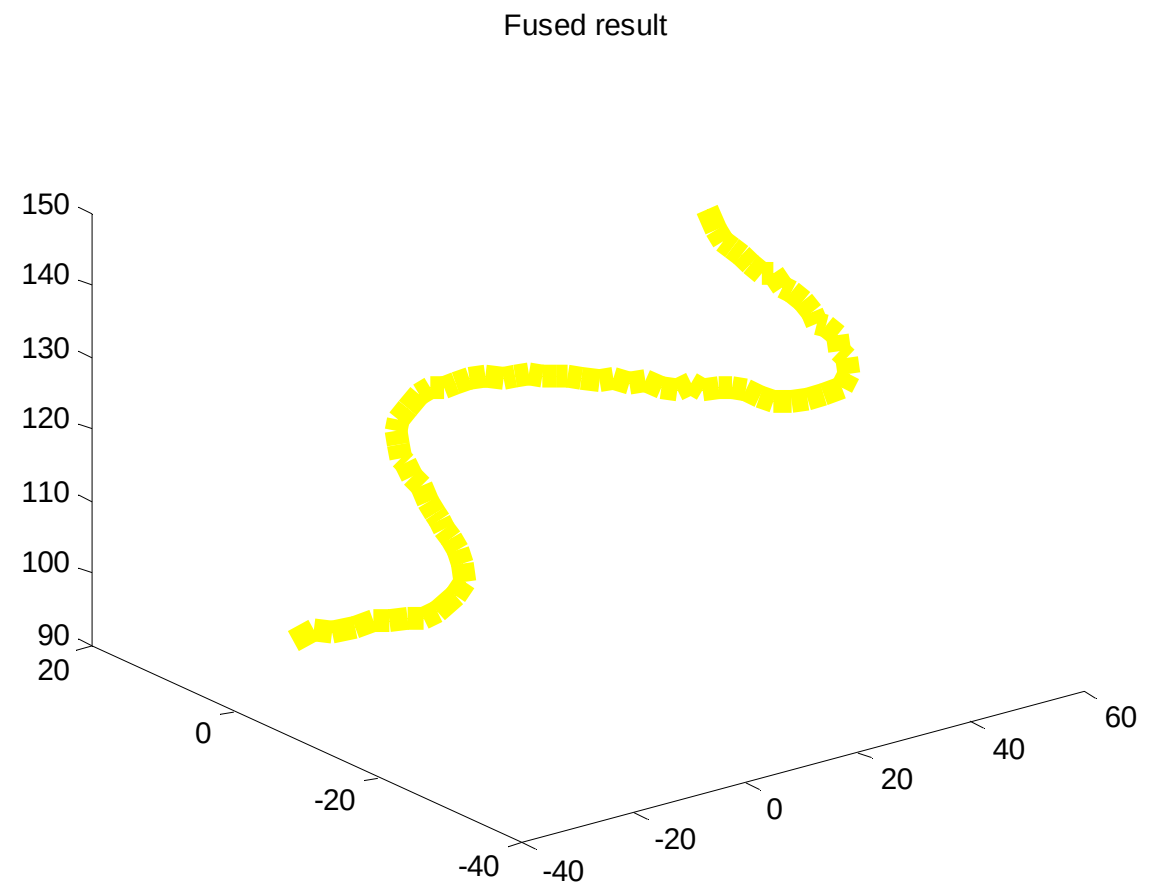

(d) Fused result

Figure 8. (a) Sensor outputs for the four sources, with added DC offset. (b) Sensor outputs for each source plotted individually. (c) Conflict measure between sources 1 and 3 plotted versus the sub-interval length. (d) Fused result.

### Example 3. Gaussian noise

In this example, 30dB Gaussian noise is added to source 2's *x, y* and *z* value, and this noise source is treated as source 3, which makes the source not as smooth as the others, which are shown in Figure 9 (a) and (b). This case mimics a sensor with a very poor SNR. This could be because of the sensor being inferior in quality to the other sensors, or due to some strong interference source. Figure 8(e) is the result using the average operator. We can see that the fused result in Figure 8(d) has reduced the effect of the Gaussian noise.

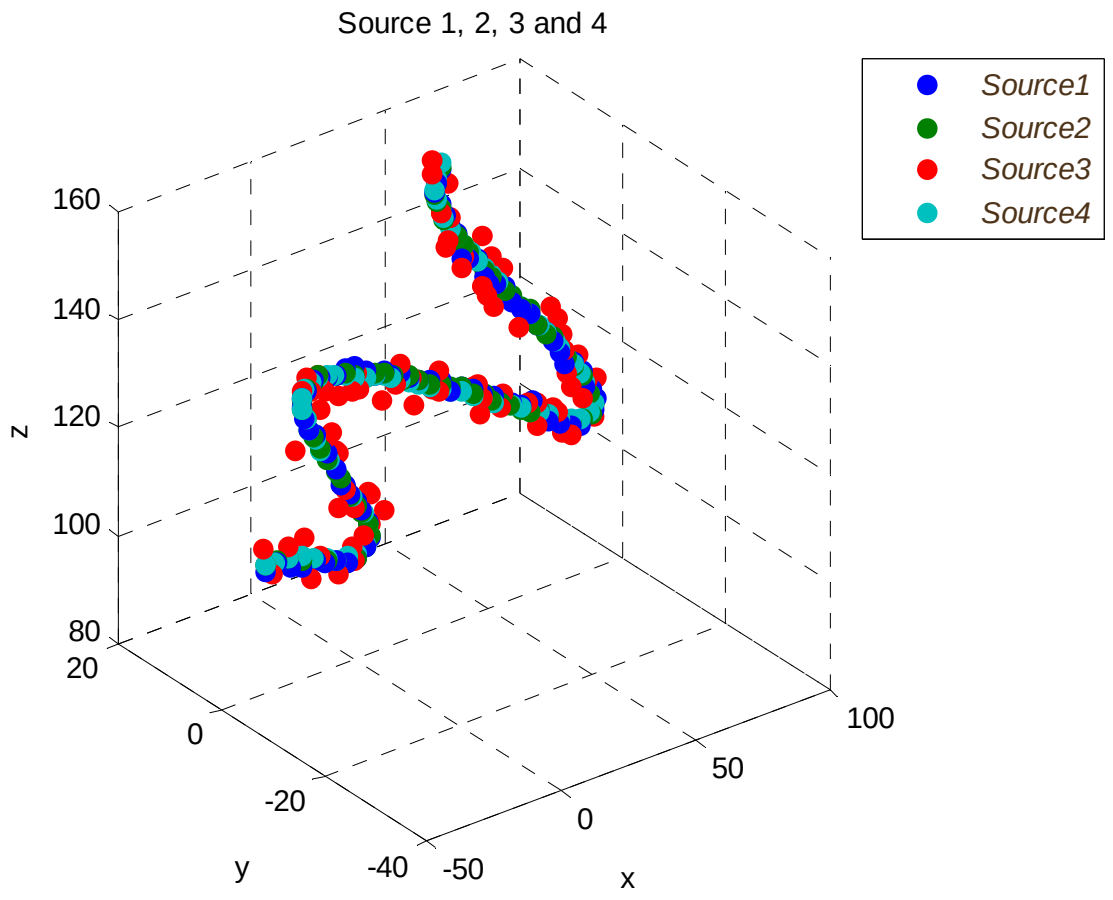

(a) Sensor outputs for all four sensors

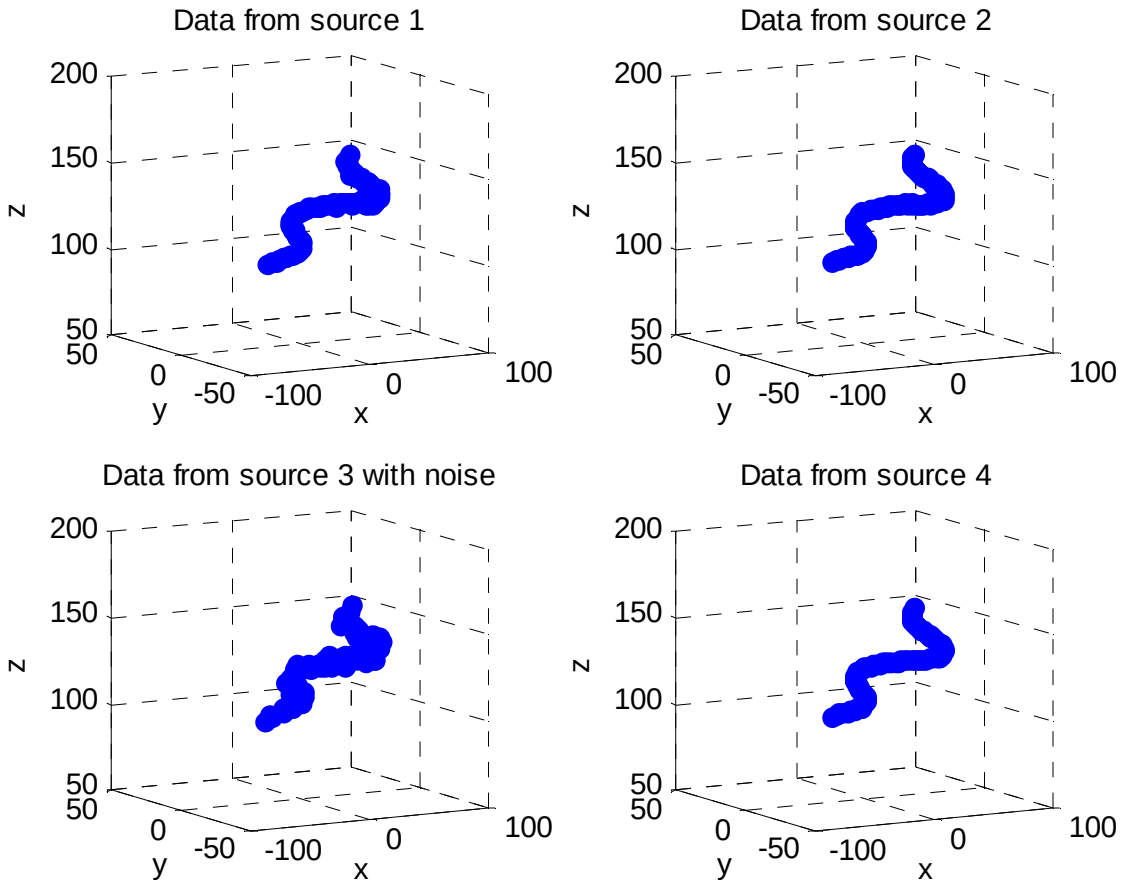

(b) Sensor outputs for each sensor

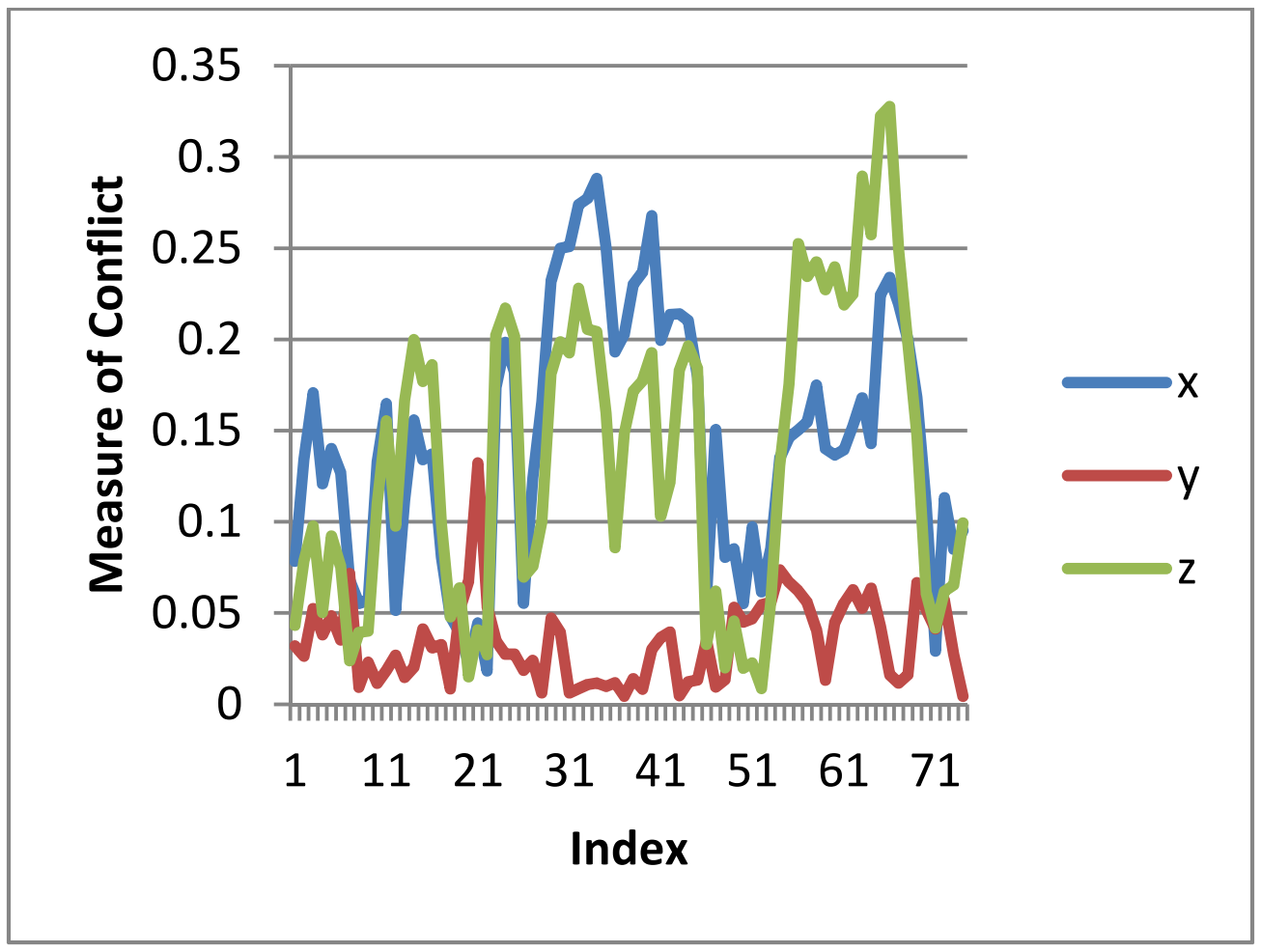

(c)Measure of Conflict of source 1 and source 3

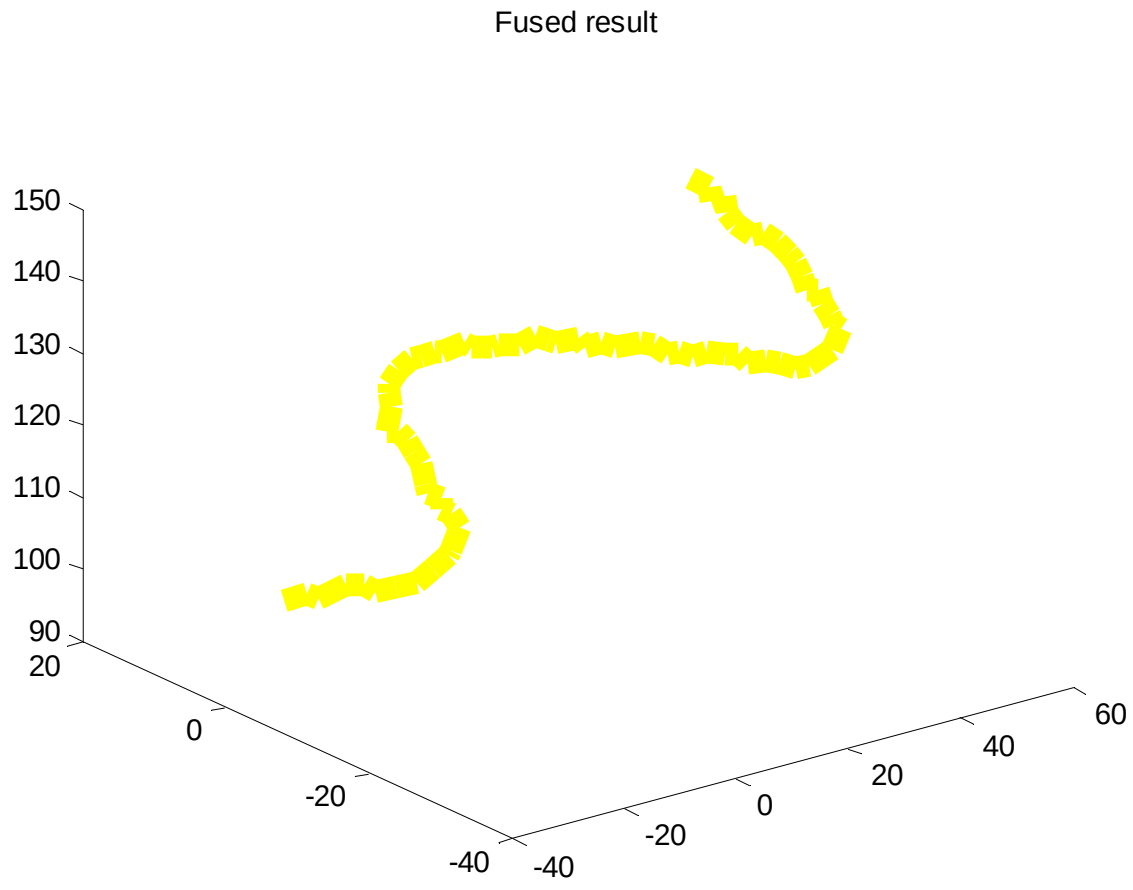

(d)  Fused result

Figure 9. (a) Sensor outputs for the four sources, with added Gaussian noise. (b) Sensor outputs for each source plotted individually. (c) Conflict measure between sources 1 and 3 plotted versus the sub-interval length. (d) Fused result.

In order to provide quantitative analysis of the results, source 1, source 2, and source 4 are the original measurements, while source 3 is generated by taking source 2 adding noise, we can suppose that source 1 is our benchmark (ground truth), and we can compare the average of absolute difference with source 1 (as shown in Equation 2) using an average operator and our proposed method. For Impulse noise and DC offset, the noise is only added to *z* values; for Gaussian noise, the noise is added to *x, y,* and *z* values. As a result, we only compare the difference in *z* between source 1 and fused result in the first two scenarios (impulsive noise and DC offset), and compare the difference in *x*, *y*, and *z* between source 1 and fused result in the last scenario (Gaussian noise) as shown in Table 1.

$$D(A, B) = \sum_{i=1}^{n} |A_i - B_i| / n \quad (6)$$

Table 1 Comparison between source 1 and the fused result using average operator or my method

| | **Average** | | | **Proposed method** | | |
|---|---|---|---|---|---|---|
| | **$D_x$** | **$D_y$** | **$D_z$** | **$D_x$** | **$D_y$** | **$D_z$** |
| **Impulse noise** | | | 0.6508 | | | **0.5392** |
| **DC offset** | | | 25.0332 | | | **4.4719** |
| **Gaussian noise** | 0.5693 | 0.2653 | 0.6162 | **0.1424** | **0.1014** | **0.5542** |

From Table 1, we can see that our method provides performance improvements versus a simple weighted sum. In the impulsive noise case, the improvement is small since there are only five impulse noise signals added. In the DC offset case, the DC offset is severe and the resulting improvement is significant. Note the x and y data in both the impulsive noise and DC offset cases were not degraded, so those entries in the table are not shown. In the Gaussian noise case, the improvements also show the diminished effects of the Gaussian noise.

## 5. CONCLUSIONS AND FUTURE WORK

In this paper, a measure of conflict using interval-valued input is proposed. By evaluating the conflict measure, we can assess credibility of sources and fuse the information from different sources together. In the fused result, the effect of the sources with higher conflict measures is diminished, while the sources with lower conflict measure play a bigger role in the final fused results. Using the proposed algorithm to fuse 3D tracking information, the results we have are better than using average operator, which gives equal credibility to every source. This shows that conflict measure can be used as an way to calculate source credibility, and the fused result based on this credibility is better than an average operator.

This work is really initial steps into this area, and more sophisticated algorithms and analysis are planned. Future work also includes the following: (1) Develop more algorithms for conflict measure estimation, and explore the mathematical characteristics of the conflict measures. (2) Fuse different conflict measures using a Fuzzy Inference System (FIS) or other algorithms. (3) Perform more experiments with additive noise of different types and levels. (4) Utilize the conflict measure at the feature level and fuse information from different features. (5) Develop rigorous, and robust optimal (or-near optimal) conflict measures for data fusion.

## 6. ACKNOWLEDGEMENT

The authors acknowledge the support of the Mississippi State University Center for Advanced Vehicular Research (CAVS) for providing the sensors and the sensor lab where this study was carried out.